\documentclass[pra,twocolumn,showpacs,nobibnotes,floatfix,superscriptaddress]{revtex4}

\usepackage{latexsym,amsmath,amssymb,amsfonts,mathbbol,graphicx,color}
\usepackage{dcolumn}
\usepackage{bm}
\usepackage{graphicx}
\usepackage{epstopdf}

\begin{document}

\title{On the Use of Shor States for the $[[7,1,3]]$ Quantum Error Correcting Code} 
\author{Yaakov S. Weinstein, Sidney D. Buchbinder}
\affiliation{Quantum Information Science Group, {\sc Mitre},
260 Industrial Way West, Eatontown, NJ 07724, USA}

\begin{abstract}
We explore the effect of Shor state construction methods on logical state encoding and quantum error correction for the [[7,1,3]] Calderbank-Shor-Steane quantum error correction 
code in a nonequiprobable error environment. We determine the optimum number of verification steps to be used in Shor state construction and whether Shor states without verification are usable for practical quantum computation. These results are compared to the same processes of encoding and error correction 
where Shor states are not used. We demonstrate that the construction of logical zero states with no first order error 
terms may not require the complete edifice of quantum fault tolerance. With respect to error correction, we show for a particular initial state that error correction using a single qubit for syndrome measurement yields a similar output state accuracy to error correction using Shor states as syndrome qubits. In addition, we demonstrate that error correction with Shor states has an inherent sensitivity to bit-flip errors. Finally, we suggest that in this type of error correction scenario one should always repeat a syndrome measurement until attaining an all zero readout (twice in row).   
\end{abstract}

\pacs{03.67.Pp, 03.67.-a, 03.67.Lx}

\maketitle

\section{Introduction}
Quantum fault tolerance \cite{G,ShorQFT,Preskill,AGP} is the framework which allows for accurate 
implementation of quantum algorithms despite the inevitability of errors during the computation. 
This is done by assuring that an error that occurs on one qubit cannot spread to multiple qubits. Application 
of quantum error correction (QEC) then corrects the single qubit error \cite{book,ShorQEC,CSS}. 

However, utilizing the entirety of the fault tolerant framework promises to be an expensive 
proposition in terms of the number of qubits and implemented gates. Thus, it is worth exploring whether 
it is possible to relax some of the strict rules required by the framework. One way to do this may 
be by easing the construction requirements or simply not using Shor states as syndrome qubits 
when encoding logical computational states and applying error correction. In this paper we study the 
utilization of Shor states in the encoding of logical zero states and the application of error correction 
for the [[7,1,3]] Steane code \cite{Steane} with the goal of limiting the number of required qubits and 
implemented gates.

A fault tolerant method for encoding a logical computational state in the Steane code is to 
apply fault tolerant error correction to any initial state of 7 qubits. This requires construction 
of proper ancilla syndrome qubits such that each ancilla interacts with no more than one of the 7 data qubits. 
For the Steane code there are a number of possible choices for these ancilla including Steane's \cite{SteaneAnc} 
suggestion of using encoded ancilla blocks, and Knill's \cite{KnillAnc} method using encoded Bell states
and teleportation. In this work we have chosen to utilize four-qubit Shor states \cite{ShorQFT} for  
ancilla as they require the least number of qubits and are thus most likely to be experimentally accessible. 
Shor states are simply Greenberger-Horne-Zeilinger (GHZ) states with Hadamard gates applied to each qubit. 
However, as the Shor states themselves are constructed in a noisy environment (here the nonequiprobable error environment), verification via parity checks is necessary to ensure accurate 
construction. In this paper, we attempt to determine the number of Shor state verifications necessary to construct logical zero states or apply error correction with as high a fidelity 
as possible. We then ask whether using Shor states with fewer verification steps (thus using fewer ancilla
qubits and requiring fewer gates) will provide sufficient accuracy to be used in the construction of 
logical zero states or the application of error correction. Finally, we explore whether Shor states 
are necessary at all in the construction of logical zeros and the application of error correction, 
or whether sufficient accuracy may be obtained using single qubits for syndrome measurement.  

The error model used in this paper is a non-equiprobable Pauli operator error model \cite{QCC} with non-correlated errors.
As in \cite{AP}, this model is a stochastic version of 
a biased noise model that can be formulated in terms of Hamiltonians coupling the system to an 
environment. In the model used here, however, the probabilities with which the different error 
types take place is left arbitrary: the environment causes qubits to undergo a $\sigma_x^j$ error 
with probability $p_x$, a $\sigma_y^j$ error with probability $p_y$, and a $\sigma_z^j$ error 
with probability $p_z$, where $\sigma_i^j$, $i = x,y,z$ are the Pauli spin 
operators on qubit $j$. We assume that only qubits taking part in a gate operation will be subject to error and 
the error is modeled to occur after (perfect) gate implementation. Qubits not involved in a gate are 
assumed to be perfectly stored. While this represents an idealization, it is partially justified in that
it is generally assumed that stored qubits are less likely to undergo error than those involved in gates
(see for example \cite{Svore}). In addition, in this paper accuracy measures are calculated only to 
second order in the error probabilities $p_i$ thus the effect of ignoring storage errors is likely minimal. 
Finally, we note that non-equiprobable errors occur in the initialization of qubits to the $|0\rangle$ state 
and measurement (in the $z$ or $x$ bases) of all qubits.

This paper builds on the previous work of Ref.~\cite{YSW} (see also \cite{BHW}) in which the fault 
tolerant method of encoding 
logical zero states for the [[7,1,3]] code was compared to the gate sequence method of encoding to see 
which method led to more accurately encoded zero states. Though the gate sequence method is not fault 
tolerant (errors can propagate to multiple data qubits) it was found that the fidelity of the logical 
zero states constructed in this way is comparable to the fidelity of the states constructed using the fault tolerant method. 
Applying perfect error correction then revealed that the error probabilities were reduced to at least
second order for both methods (third order for the fault tolerant method), implying the correctability 
of the errors and suggesting that either method can be used for practical quantum computation. Here we work 
within the fault tolerant method in attempt to determine how best to construct Shor states for encoding 
and error correction. In both papers, however, a major goal is to determine whether accurate enough protocols
can be implemented without invoking the full framework of quantum fault tolerance. 

\section{Constructing Shor States}

\begin{figure}
\includegraphics[width=6.5cm]{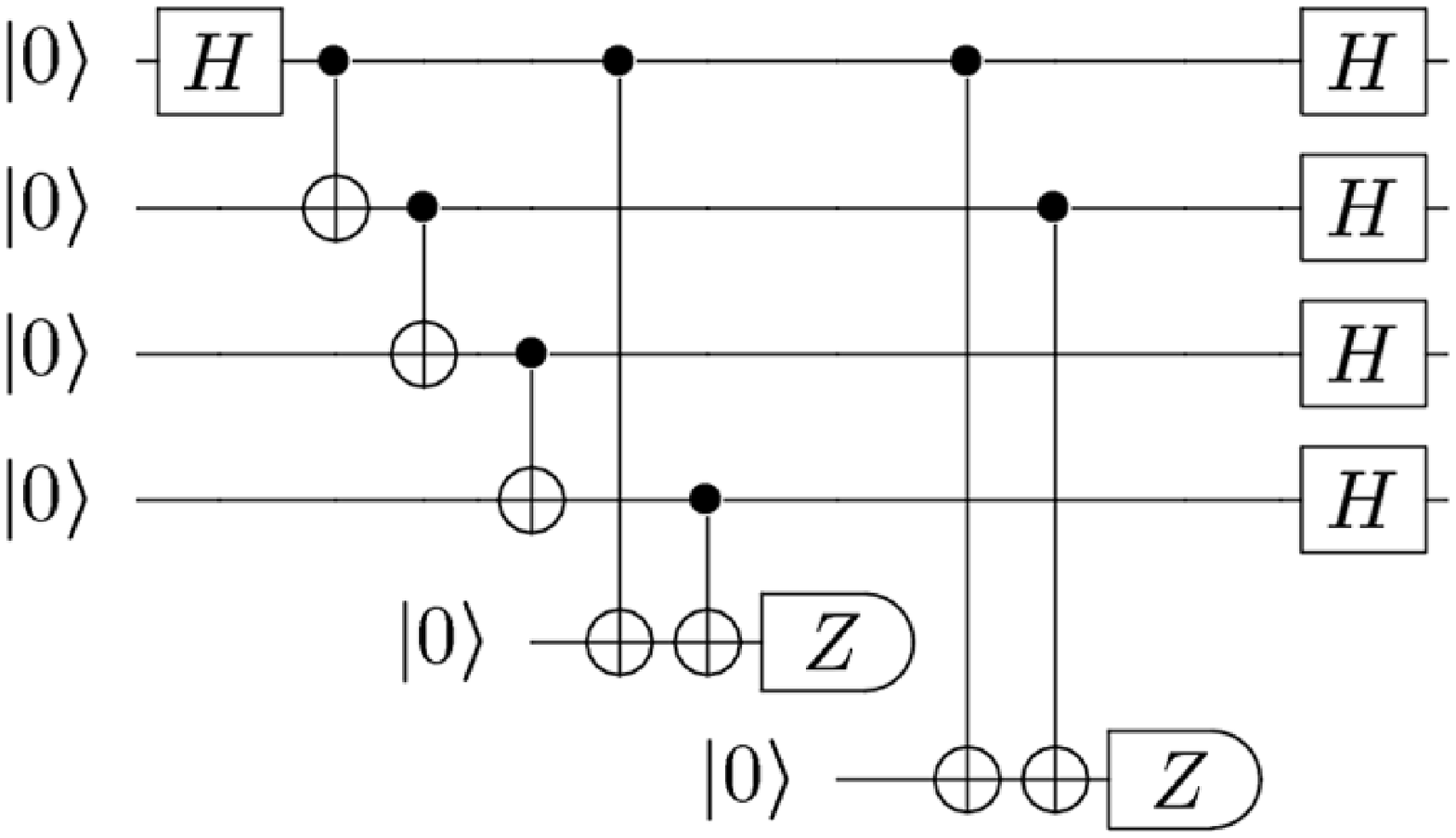}
\includegraphics[width=8.5cm]{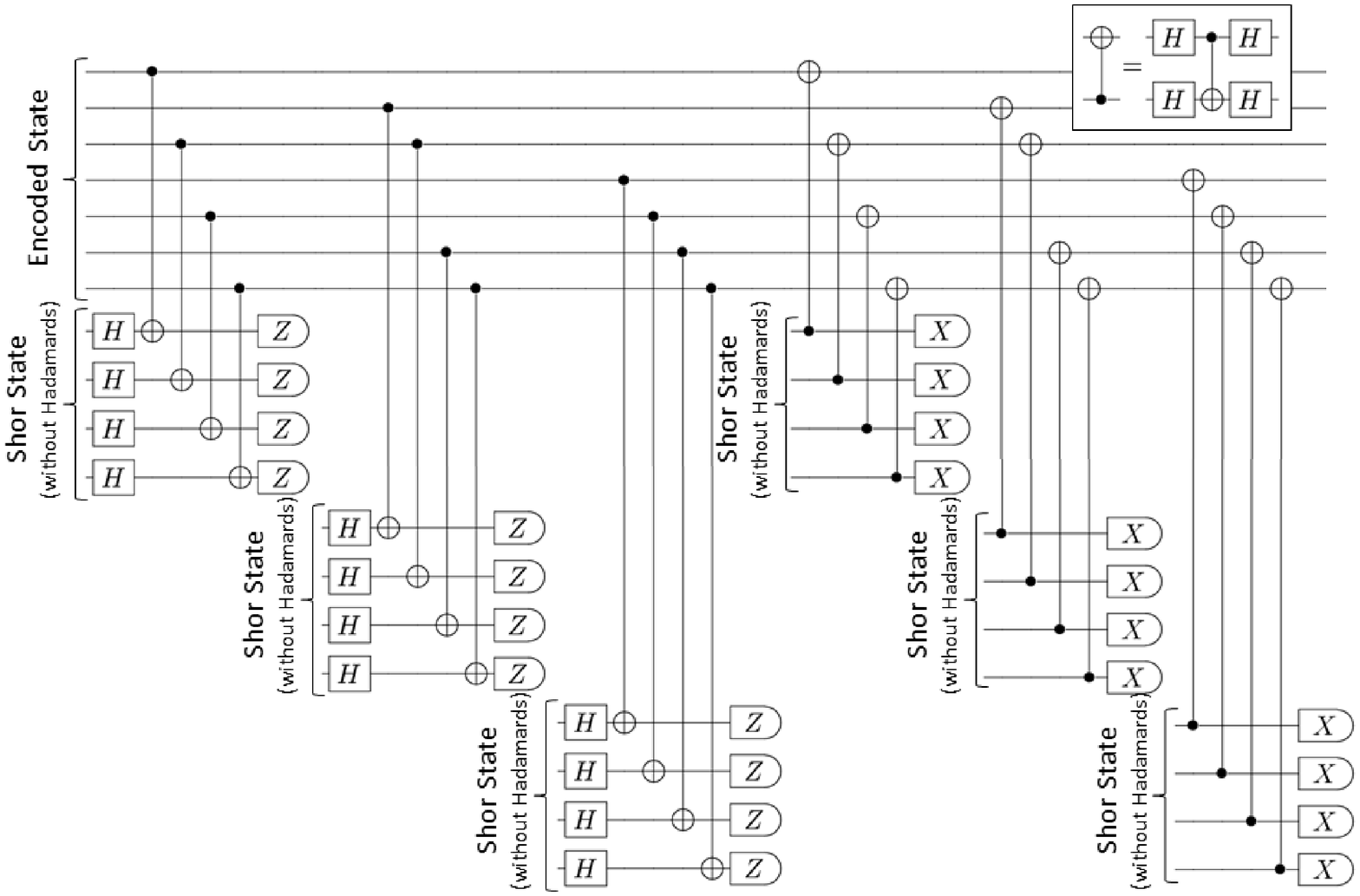}
\caption{Top: construction of a 4 qubit Shor state. \textsc{cnot} gates are represented by ($\bullet$) on the control qubit and ($\oplus$) on the target qubit connected by a vertical line. $H$ represents a Hadamard gate. The procedure entails constructing a GHZ state which is verified using ancilla qubits. Hadamard gates are applied to each qubit to complete Shor state construction.
Bottom: fault tolerant bit-flip and phase-flip syndrome measurements for the [[7,1,3]] code using Shor states (the Shor states pictured are assumed to have not had the Hadamard gates applied). To ensure fault tolerance, each Shor state ancilla qubit must interact with only one data qubit. The error syndrome is determined from the parity of the measurement outcomes of the Shor state ancilla qubits. To achieve fault tolerance each of the syndrome measurements is repeated twice. 
Box: a useful equality which allows us to avoid implementing Hadamard gates by reversing the control and target of \textsc{cnot} gates. In our context, the \textsc{cnot} gates associated with the phase-flip syndrome measurements are reversed such that the ancilla qubits become the control and the data qubits become the target, as explained in the text. }
\label{ShorState}
\end{figure}
  
A construction method for the four-qubit Shor states needed for the [[7,1,3]] QEC code 
is shown in Fig.~\ref{ShorState}. If the construction was done without error,
no verification steps would be needed and the Shor state (without the 
final Hadamard gates as explained below) would be given by:
$|\psi_{Shor}\rangle = \frac{1}{\sqrt{2}}(|0000\rangle + |1111\rangle)$. However, actual
implementations of quantum computation will be done in a noisy environment and thus verifications 
may be useful. We simulate construction of Shor states in the nonequiprobable error environment 
including initialization and measurement errors with different verification strategies. We then 
determine which of the strategies produce the highest quality Shor states based on the fidelity 
of the constructed Shor states, the fidelity of logical zero states encoded fault tolerantly with 
the different Shor states used as syndrome qubits, and the fidelity of a state after noisy error correction
when the different Shor states are used as syndrome qubits. The different strategies we 
use are: no verifcation steps, one verification step, and different possible two verification steps. 
The tenets of fault tolerance require that at 
least one verification step be applied so as to lower the probability of error to second order. 

\begingroup
\squeezetable
\begin{table*}
\caption{Relevant fidelity measures for Shor states and encoded logical zeros from different construction 
methods: Shor state without 
verification, Shor state with one verification, Shor state with two verifications, and the accuracy of 
logical zero construction using single-qubit ancilla for syndrome measurements instead of Shor states. 
The accuracy measures are the fidelity of the Shor state itself, the fidelity of the seven physical qubits 
making up the logical zero state, the fidelity of the one qubit of information stored in the seven physical
qubits, and the fidelity after perfect error corrction has been applied to the constructed encoded zero states. }
\begin{tabular}{||c||c|c|c|c||}
\hline 
 & no verifications & 1 verification & 2 verifications & 1-Qubit ancilla \\\hline
\hline
Shor fidelity & $1-10p_x-11p_y-7p_z$ & $1-5p_x-6p_y-10p_z$ & $1-5p_x-6p_y-13p_z$ & \\\hline
7-Qubit fidelity  & $1-85p_x-37p_y-12p_z$  & $1-55p_x-19p_y-12p_z$ & $1-55p_x-19p_y-12p_z$ & $1-49p_x-19p_y-12p_z$ \\\hline
1-Qubit fidelity  & $1-25p_x-11p_y$ & $1-19p_x-7p_y$ & $1-19p_x-7p_y$ & $1-15p_x-7p_y$ \\\hline
after QEC	& $1-92p_x^2-74p_xp_y-14p_y^2$ & $1$ & $1$ & $1-26p_x^2-6p_xp_y$ \\
\hline
\end{tabular}
\label{Tab1}
\end{table*}
\endgroup

To construct the Shor state we start with four qubits that we attempt to initialize to the state zero. 
However, in this work, we assume that initialization itself is a noisy process subject to the same 
error model as qubits involved in a gate. Thus, the actual state of each initialized qubit is 
$\rho_i=(1-p_x-p_y)|0\rangle\langle 0|+(p_x+p_y)|1\rangle\langle 1|$. We then apply a Hadamard 
gate, $H$, to the first qubit. The nonequiprobable error environment causes imperfections in the gate such 
that the actual evolution of an attempted Hadamard on a single qubit $j$ in the state $\rho$ is: 
\begin{equation}
\sum_{a=0,x,y,z}p_a\sigma_a^jH_j\rho H_j^{\dag}\sigma_a^j,
\end{equation} 
where $\sigma_0^j$ is the identity matrix, $p_0 = 1-\sum_{\ell=x,y,z}p_\ell$,
and the terms $K_a^j = \sqrt{p_a}\sigma_a^jH_j$ can be regarded as Kraus operators for the 
Hadamard evolution. The Hadamard is followed by a series of \textsc{cnot} gates. The attempted 
performance of the \textsc{cnot} gate with control qubit $j$ and target qubit $k$, $\textsc{c}_j\textsc{not}_k$,
in the nonequiprobable error environment on any state $\rho$ actually implements: 
\begin{equation}
\sum_{a,b=0,x,y,z}p_ap_b\sigma_a^j\sigma_b^k\textsc{c}_j\textsc{not}_k\rho \textsc{c}_j\textsc{not}_k^{\dag}\sigma_a^j\sigma_b^k,
\label{cnot}
\end{equation}
where terms $A_{a,b}^{j,k} = \sqrt{p_ap_b}\sigma_a^j\sigma_b^k\textsc{c}_j\textsc{not}_k$ can be regarded as the 16 Kraus 
operators. Note that errors on the two qubits taking part in the CNOT gate are independent and not correlated. Shor state construction requires three \textsc{cnot} gates, shown in Fig.~\ref{ShorState}, and thus 
the final Shor state is given by
\begin{eqnarray}  
\rho_{Shor-err} &=& \sum_{a,b,c,d,e,f,g}^{0,x,y,z}A_{f,g}^{3,4}A_{d,e}^{2,3}A_{b,c}^{1,2}K_a^1\rho_i^{\otimes 4} \nonumber\\
		    &\times& (K_a^1)^\dag(A_{b,c}^{1,2})^\dag(A_{d,e}^{2,3})^\dag(A_{f,g}^{3,4})^\dag.
\end{eqnarray}

As explained above, applying the above described gate sequence does not guarantee that the resulting Shor states are suitable for fault tolerant quantum computation. During syndrome measurement, errors in the Shor state construction can 
propagate into the data qubits. If only one Shor state qubit has been compromised
by error then only one data qubit will be compromised and the error can be subsequently
corrected. However, if multiple Shor state qubits are compromised, more than one data
qubit can be compromised and the computation will fail. Thus, we must test the Shor states 
to ensure that multiple qubits have not been compromised by error. 
This is done utilizing an ancilla qubit, initially in the 
state $|0\rangle$, adjoined to the Shor state to measure the parity of random pairs of 
qubits \cite{ShorQFT}. Should the test fail (the ancilla qubit measurement yields a 
$|1\rangle$), the Shor state is immediately discarded. Of course, the ancilla qubit 
initialization and the \textsc{cnot} gate implementations
for this parity check are themselves performed in the nonequiprobable
error environment and thus follow the dynamics described above. 
We utilize an initial ancilla qubit to measure the parity of qubits 1 and 4.
Applying additional verification steps using additional ancilla may, if the \textsc{cnot}s themselves
are not too error prone, further ensure the lack of errors in the constructed Shor states. 
A second ancilla can recheck the parity of the qubits checked with the first ancilla, or check the parity between other Shor state qubits. We have simulated every possible combination 
for the second parity measurement and have found that this choice has little effect on any of our accuracy measures.

Our first accuracy measure for the Shor states constructed with different numbers of verifications
is the fidelity of the constructed Shor state as compared to a perfect Shor state, 
$F = \langle\psi_{Shor}|\rho_{Shor-err}|\psi_{Shor}\rangle$. The fidelity results for 
Shor states with zero, one, and two parity verifications are shown to first order in 
error probability in the first row of Table \ref{Tab1}. Note that to first order the fidelity 
for Shor states of two parity verifications is independent of which qubits are used for 
the second verification. 

Comparing the fidelity of the three Shor states we see that the Shor state with one verification
has a higher fidelity than the Shor state with no verifications unless $p_z$ is significantly 
higher than $p_x$ and $p_y$. This demonstrates that it is usually advisable to perform a verification 
step in order to suppress errors that occur during the Shor state construction. However, the fidelity of 
the Shor state with two verification steps is always lower than that of the Shor state with one verification
step. A second verification step does not give enough benefit to outweigh additional errors that may occur 
during the verification procedure.

\section{Encoding with Shor States} 

The Shor state fidelity is a good measure of accuracy for the Shor state in and of itself.
However, our purpose for constructing Shor states is to use them to encode logical
zero states and implement fault tolerant error correction. It is possible that different errors in 
the Shor state construction will have more or less of an effect on the accuracy with which 
these protocols can be performed. Thus, another way to quantify the quality of the Shor states 
is to simulate their utilization in the encoding of logical zero states 
and in the performance of error correction and report on the accuracy with which these 
protocols are implemented. 

We first turn to the construction of logical zero states. To do this in a fault tolerant manner
we start with 7 qubits all noisily initialized to the state zero. Though this initialization is not perfect 
we choose to not perform the first set of (bit-flip) syndrome measurements as their utility in 
correcting an initialization error is outweighed by the noise inherent in applying the necessary 
syndrome measurments. 
Instead, we immediately measure the three phase flip syndromes (each one of the three twice) with Shor states as the syndrome qubits. To measure phase flip syndromes requires applying a Hadamard gate to each of the seven data qubits before and after the syndrome measurements. However, we can measure the syndrome without Hadamard gates if we reverse the roles of the control and target qubits for the \textsc{cnot} gates, and measure the Shor state qubits (noisily) in the $x$-basis, as explained in 
\cite{Preskill} and shown in Fig.~\ref{ShorState}. For the case of encoding we analyze the scenario where all syndrome results are zero. Because
encoding is done `off-line' one can choose to utilize only the encoded states with this outcome. 

Encoding in the nonequiprobable error environment using Shor states with different 
numbers of applied verifications will result
in logical zero states with different degrees of accuracy. We can measure this accuracy in a number of ways. 
The first way is simply to look at the fidelity of the seven qubit logical zero state. The accuracy of 
this state gives an idea as to how well the entire encoding process was performed. Alternatively, 
one may look at the fidelity of only the one qubit of encoded information. This is the only qubit 
of information that is actually of importance and, if it is protected, the state of the rest of the system 
is irrelevant. Measuring the fidelity of this one logical qubit is done by (noiselessly) decoding the 
constructed logical zero state, tracing out all qubits but the first, and comparing the state 
of the remaining qubit with the zero state on a single qubit. Both of these fidelity 
measures have been calculated for logical zero states constructed using Shor states of zero, one,
and two verification parity checks, and are given in Table \ref{Tab1}.

Errors affecting the logical zero state may also be of varying degrees of severity. Applying perfect 
error correction allows us to test the `correctability' of the types of 
errors that occur during the encoding. If even perfect error correction cannot (to first order) 
correct the errors in the logical zero state then the encoding method cannot be 
used for practical implementations of quantum computation. We apply perfect error correction to 
the states constructed using the Shor states with varying numbers of verifications
and calculate the fidelity measure of the output state. These fidelities are given in Table \ref{Tab1},
and corroborate our previous observations that applying one verifiction to Shor states is optimal. 
Applying no verification steps to the Shor states leads to lower fidelities for the 
logical zero states, and applying two verifications does not raise the fidelity. Perfect error 
correction applied to logical zero states encoded using Shor states with one or two verifications 
gives unit fidelity up to third order. However, perfect error correction applied to logical zero 
states encoded using Shor states with no verifications, suppresses errors to second order implying
that these states may also be useable for practical quantum computation.  

We compare the above cases of Shor state syndrome measurement with a logical zero encoding method 
in which a single ancilla qubit is used for each syndrome measurement. This method does not meet the 
standards of fault tolerance since an error on the single ancilla qubit, be it an initialization error or an 
error in one of the syndrome measurement \textsc{cnot} gates, can spread to multiple data qubits. However, 
using one ancilla qubit removes the need to construct Shor states thus lowering the 
number of gates to be performed. The logical zero fidelity measures defined above are calculated for the single
qubit syndrome measurement construction method and are shown in Table \ref{Tab1}. Comparing these fidelity 
measures to those calculated for Shor state based encoding, we find that using single qubit ancilla leads to higher 
fidelity logical zero states. However, upon application of perfect 
error correction the error probabilities are suppressed only to second order, unlike the logical zero states 
constructed using Shor states for which the second order error probability terms are also suppressed.
  
\section{Quantum Error Correction with Shor States}

We now consider the accuracy with which the different Shor states can be used as syndrome ancilla qubits 
for quantum error correction. The arbitrary single-qubit initial state we would like to protect is assumed 
to have been perfectly encoded via the [[7,1,3]] gate encoding sequence: 
$|\psi\rangle=\cos\alpha|0_L\rangle+e^{i\beta}\sin\alpha|1_L\rangle$, where 
$|0_L\rangle$ and $|1_L\rangle$ represent the seven qubit logical $|0\rangle$ and $|1\rangle$ states
respectively. We assume the environment possibly causes an error such that, before error correction, 
the system is in a mixed state of no error and all possible single qubit errors: 
\begin{equation}
\rho_{err}=(1-7(p_x+p_y+p_z))|\psi\rangle\langle\psi|+\displaystyle\sum\limits_{i=1}^7\sum\limits_{a=x,y,z} p_a\sigma_a ^i |\psi\rangle\langle\psi|{\sigma_a^i}^{\dagger}.
\end{equation}
Because there are only single qubit errors in the system state, the error can be corrected by perfect 
application of the [[7,1,3]] code. 

\begingroup
\squeezetable
\begin{table*}
\caption{Fidelity measures for error correction applied to the state $\rho_{err}$ utilizing Shor states with different 
numbers of verifications or a single ancilla qubit for syndrome measurement. In the Table $a = \cos[4\alpha]$ and 
$b = \cos[2\beta]\sin[2\alpha]^2$. In this case the bit flip syndrome measurements were done first. }
\begin{tabular}{||c||c|c|c|c||}
\hline 
 & no verifications & 1 verification & 2 verifications & 1-Qubit ancilla \\\hline
\hline
7-Qubit fidelity  & $1-85p_x-25p_y-7p_z$ & $1-55p_x-7p_y-7p_z$ & $1-55p_x-7p_y-7p_z$ & $1-49p_x-7p_y-7p_z$ \\\hline
			& $1-(\frac{81}{4}+\frac{27}{4}a-\frac{27}{2}b)p_x$ & $1-(\frac{57}{4}+\frac{19}{4}a-\frac{19}{2}b)p_x$ & $1-(\frac{57}{4}+\frac{19}{4}a-\frac{19}{2}b)p_x$ & $1-(\frac{45}{4}+\frac{15}{4}a-\frac{15}{2}b)p_x$ \\
1-Qubit fidelity  & $-(\frac{25}{4}+\frac{3}{4}a-\frac{5}{2}b)p_y$ & $-(\frac{13}{4}-\frac{1}{4}a-\frac{1}{2}b)p_y$ & $-(\frac{13}{4}-\frac{1}{4}a-\frac{1}{2}b)p_y$ & $-(\frac{13}{4}-\frac{1}{4}a-\frac{1}{2}b)p_y$ \\
			& $-\frac{3}{2}(1-a)p_z$ & $-\frac{3}{2}(1-a)p_z$ & $-\frac{3}{2}(1-a)p_z$ & $-\frac{3}{2}(1-a)p_z$ \\\hline
\end{tabular}
\label{Tab2}
\end{table*}
\endgroup

To perform error correction in a fault tolerant manner, Shor states with at least one verification must 
be used for syndrome measurements. We apply error correction to the state $\rho_{err}$ in the nonequiprobable
error environment by implementing 
the three bit-flip syndrome measurements followed by three phase-flip syndrome measurements using Shor 
states with different numbers of verifications as the syndrome qubits. Each syndrome 
measurement is repeated twice to account for errors that may have occurred during the syndrome measurement 
itself. We quantify the quality of the error correction via fidelity measures comparing the final state 
after error correction to the pre-encoded arbitrary state. 

\subsection{Syndrome Measruement Reveals No Error}

To read out the syndrome bit the four Shor state qubits are measured. If the results of the four measurements are of even parity the syndrome bit is a zero. If the results are of odd parity the syndrome bit is a one. We first look at the case where all qubit measurements are zero. Other even parity measurment results (say 0011 or 0101) give the same fidelity to first order. The fidelities of the seven data qubits and the one logical qubit state for this case are given in Table \ref{Tab2}. 

Comparing the seven-qubit fidelities of the QEC procedure utilizing Shor states with different numbers of verifications, we first note that the fidelities 
for Shor states with one and two verifications are identical up to second order terms. This fidelity is higher 
than that attained by performing QEC using a Shor state with no verifications, again confirming that while 
performing verification of the Shor state is important, there is no benefit gained from performing a second verification step. The fidelities exhibit little dependence on the initial state of the qubit, $\alpha$ and $\beta$ only appear in second order fidelity terms. Furthermore, regardless of the Shor state used, the $p_x$ error is dominant implying that 
bit-flips are more harmful to the error correction procedure than phase flips. 

Similar trends hold when comparing the single qubit fidelities except that all of the single 
qubit fidelities depend strongly on the initial state. These fidelities are highest when $\alpha=0,\frac{\pi}{2}$, at which point the first and second order $p_z$ terms drop from the fidelity expression, and are lowest when $\alpha=\frac{\pi}{4}$. 
Once again $p_x$ is the dominant error term. We note that the presence of first order terms 
in the fidelity measures indicate that, in this case, noisy QEC cannot output a state with 
no first-order error probability terms. Practical quantum error correction in this case is 
thus reduced to minimizing the coefficients of these first order terms. 

We compare the above QEC performance with that of error correction done without Shor states, instead using a single (noisily initialized) ancilla qubit for syndrome measurement. While this scheme certainly does not meet the criteria for fault tolerance, it does allow us to implement QEC with fewer qubits, and the lack of possible error from the construction of the Shor states may, and in fact does, yield an improved resulting fidelity. The fidelities for this case are shown in the last column of Table \ref{Tab2}.

\subsection{Syndrome Measruement Reveals Error}

Above we assumed that errors occur with low probability ($p_j \ll 1$) and thus the chances of measuring a bit-flip or phase-flip syndrome that is not 000 is extremely small. If, however, syndrome measurement does (twice in a row) signify an error a proper recovery operation must be performed. In such a case we find extremely low fidelities ($\approx .5$). The explanation for this is as follows (referencing Fig. 1, this should be compared to the fidelity results of \cite{BHW}): let us say that the syndrome 001 is measured (twice in a row) presumably indicating that the fourth qubit has undergone an error. Note, that the same syndrome would arise if an error had occurred to, say, qubit 7 during or after the $\textsc{c}_7\textsc{not}_{11}$ gate of the second syndrome bit. If this latter error had occurred the recovery operation would be applied to the wrong qubit, and, thus, the final state would have two errors: the error on qubit 7 which was not corrected and the error on qubit 4 due to the mistaken recovery operation. Because both the gates associated with error correction and the gates applied before error correction are implemented in the same error environment, there is no \emph{a priori} reason to think that this latter scenario is any less probable then the presumed error based on the syndrome measurement. Therefore, the final state of the system after the error correction procedure is a mixed state consisting of a corrected state and states with two errors (not to mention terms from errors that may have occurred during the final syndrome measurement, recovery operation, etc.) leading to an unacceptably low fidelity. 

The above suggests a proper procedure to follow upon obtaining a non-zero syndrome measurement (even if the same syndrome is read out twice in a row). Rather then accepting the syndrome measurement and applying the requisite recovery operation, the syndrome measurement should be redone until the all zero readout is attained (twice in a row). Proceeding based on the non-zero readout will likely lead to a state with uncorrectable errors.  

\subsection{Why Bit Flips?}

We noted above that $\sigma_x$ errors dominate the loss of fidelity. There are a couple of possibilities as to why this may be so. The first is because the bit-flip syndrome measurements were implemented first, and thus $\sigma_x$ errors that may occur during phase-flip syndrome measurements are not corrected.
A second possibility is that the use of (noisy) Shor states may cause the effect of $\sigma_x$ errors to be more 
pronounced. In this section we clarify this issue by carrying out a series of simulations designed to isolate 
the cause of increased sensitivity to $\sigma_x$ errors. 
 
Our first step is to repeat the above error correction calculations implementing the phase-flip syndrome 
measurements first. The fidelities of the resulting states are shown in Table \ref{Tab3}.
Let us first compare the cases where the syndrome measurement was done with a single ancilla qubit.
In this case the coefficients of the $p_x$ and $p_z$ terms in the seven qubit fidelity simply switch places 
while the $p_y$ coefficient remains constant. Similarly in the one-qubit fidelity the $p_y$ coefficient 
remains constant while the values of the $p_x$ and $p_z$ terms approximately trade values (modulo the
contribution of the initial state). This alone suggests that the dominance of the $p_x$ term
in the original simulations was simply because the bit-flip syndrome measurements were done first. 
When the phase-flip syndrome measurements are done first $p_z$ replaces $p_x$.  
However, when looking at the QEC simulations that utilize Shor states for syndrome measurements 
we do not find the same trade-off. Instead, though the $p_z$ error coefficients
grow and (in most cases) become dominant, we find much less of a reduction of the 
$p_x$ error coefficients. This suggests that there is something inherent in the use of the 
(noisy) Shor states that leads to this type of error. 

\begingroup
\squeezetable
\begin{table*}
\caption{Fidelity measures for error correction applied to the state $\rho_{err}$ utilizing Shor states with different 
numbers of verifications or a single ancilla qubit for syndrome measurement. In the Table $a = \cos[4\alpha]$ and 
$b = \cos[2\beta]\sin[2\alpha]^2$. In this case the phase flip syndrome measurements were done first. }
\label{fidtable}
\begin{tabular}{||c||c|c|c|c||}
\hline 
 & no verifications & 1 verification & 2 verifications & 1-Qubit ancilla \\\hline
\hline
7-Qubit fidelity  & $1-61p_x-25p_y-55p_z$ & $1-31p_x-7p_y-55p_z$ & $1-31p_x-7p_y-55p_z$ & $1-7p_x-7p_y-49p_z$ \\\hline
			& $1-(\frac{61}{4}-\frac{49}{4}a-\frac{3}{2}b)p_x$ & $1-(\frac{33}{4}-\frac{21}{4}a-\frac{21}{2}b)p_x$ & $1-(\frac{33}{4}-\frac{21}{4}a-\frac{3}{2}b)p_x$ & $1-(\frac{9}{4}+\frac{3}{4}a-\frac{3}{2}b)p_x$ \\
1-Qubit fidelity  & $-(\frac{33}{4}-\frac{21}{4}a-\frac{1}{2}b)p_y$ & $-(\frac{13}{4}-\frac{1}{4}a-\frac{1}{2}b)p_y$ & $-(\frac{13}{4}-\frac{1}{4}a-\frac{1}{2}b)p_y$ & $(\frac{13}{4}-\frac{1}{4}a-\frac{1}{2}b)p_y$ \\
			& $-\frac{27}{2}(1-a)p_z$ & $-\frac{27}{2}(1-a)p_z$ & $-\frac{27}{2}(1-a)p_z$ & $-\frac{25}{2}(1-a)p_z$ \\\hline
\end{tabular}
\label{Tab3}
\end{table*}
\endgroup

To further explore this point we perform two additional sets of QEC simulations. In the first,
we utilize perfect Shor states but allow errors (due to the nonequiprobable error environment) in 
the error correction process (including syndrome measurement). In the second, we use Shor states
constructed in the nonequiprobable error environment (with one verification) but the error 
correction itself (including syndrome measurements) is perfect. Both are done with bit-flip 
syndrome measurements first and with phase-flip syndrome measurements first. When perfect 
Shor states are used, but the error correction is noisy, we find that the dominant error 
depends on which set of syndrome measurements is done first, if phase correction is done
first $\sigma_z$ errors dominate and vice-versa. The other error type is significantly diminished.
When noisy Shor states are used with perfectly implemented error correction we find that which 
syndrome is done first makes little difference: $\sigma_x$ errors dominate and the fidelities do not
contain a first order term for $\sigma_z$ errors. The various fidelity measures are displayed in 
Table \ref{Tab4}.

\begingroup
\squeezetable
\begin{table*}
\caption{Fidelity measures for quantum error correction applied to the state $\rho_{err}$ with perfect Shor 
states and noisy error correction, and noisy Shor states with perfect error correction. Both cases were done 
with the $\sigma_x$ syndrome measurements first and the $\sigma_z$ syndrome measurements first. In the Table 
$a = \cos[4\alpha]$ and $c = \cos[2\beta]$.}
\begin{tabular}{||c||c||c|c||}
\hline 
 & & bit-flip first & phase-flip first \\\hline
\hline
				& 7-Qubit fidelity	& $1-31p_x-7p_y-7p_z$ & $1-7p_x-7p_y-55p_z$ \\\cline{2-4}
Noisy QEC			& 				& $1-(\frac{33}{4}+\frac{11}{4}(a-c+ac)p_x$ & $1-(\frac{9}{4}+\frac{3}{4}a-\frac{3}{2}b)p_x $ \\
Perfect Shor States	& 1-Qubit fidelity  	& $-(\frac{13}{4}-\frac{1}{4}(a+c-ac)p_y$ & $-(\frac{13}{4}-\frac{1}{4}a-\frac{1}{2}b)p_y$ \\
				&				& $-\frac{3}{2}(1-a)p_z$ & $-\frac{27}{2}(1-a)p_z$ \\\hline
Perfect QEC			& 7-Qubit fidelity	& $1-24p_x$ & $1-24p_x$ \\\cline{2-4}
Noisy Shor States		& 1-Qubit fidelity	& $1-(6+2a-4b)p_x$ & $1-(6+6a)p_x $ \\\hline
\end{tabular}
\label{Tab4}
\end{table*}
\endgroup

Taken together these simulations imply that when noisy Shor states are utilized for syndrome measurement
in the Steane code there is a significant bias towards bit-flip errors. A possible solution is to  
concatenate into a three-qubit bit-flip QEC code for another level of error correction. This could 
significantly reduce the sensitivity to bit-flip errors without the resource cost of concatenation into 
another level of the seven-qubit Steane code.

\section{Conclusion}

In conclusion, we have calculated quality metrics for different Shor states used as syndrome
measurement ancilla qubits for the [[7,1,3]] CSS QEC code operating in a nonequiprobable error 
environment. The results suggest that while a Shor state constructed in this error environment 
with one parity check verification is optimal for suppressing errors in the construction of 
logical zero states, Shor states with no checks will also suppress error probability terms in the fidelity to second order. In addition, encoding applied without Shor states, instead using single qubit ancilla for syndrome measurement, leads to logical zero states with
higher fidelity but errors that are less correctable as identified by fidelity after perfect 
error correction. 

For error correction applied in a nonequiprobable error environment using the seven qubit Steane code, 
our simulations show that not using Shor states leads to a corrected state with higher fidelity than using Shor states. In addition, we noted that bit-flip errors are dominant whether
Shor states are used or not. We first suggested that this was due to the fact that the bit-flip syndrome measurements were 
done first, meaning that uncorrected bit-flips may accumulate during phase-flip syndrome measurements.
Simulations switching the order of the syndrome measurements demonstrated that this is correct when 
using single qubit ancillae for syndrome measurement, but does not completely explain the results of
simulations using Shor
states. Further simulations indicated an inherent sensitivity towards bit-flip errors when Shor states
are used. We suggested that this could be overcome by concatenating with a three-qubit QEC code
that protects against bit-flip errors. Finally, we suggested that when a non-zero syndrome is detected implementing the prescribed recovery operation will lead to a state of unacceptably low fidelity. Rather the syndrome measurement should be repeated until a zero syndrome readout is attained.

The authors would like to thank G. Gilbert for constructive comments. This research is supported under MITRE Innovation Program Grant 07MSR205.

\end{document}